\documentclass[doublecol]{epl2} 

\usepackage{amssymb}

\title{Surface-charge-induced freezing of colloidal suspensions}
\shorttitle{Surface-charge-induced freezing of colloidal suspensions} 

\author{S.~Grandner\inst{1} \and S.~H.~L.~Klapp\inst{1}}
\shortauthor{S.~Grandner \etal}

\institute{                    
  \inst{1} Institut f\"ur Theoretische Physik, Technische Universit\"at Berlin - Hardenbergstra{\ss}e 36, D-10623 Berlin, Germany
}

\newcommand{\av}[1]{\left\langle #1 \right\rangle}
\newcommand{\kB}{k_\mathrm{B}}

\abstract{
Using grand-canonical Monte Carlo simulations we investigate the impact of charged walls on the crystallization properties of charged colloidal suspensions confined between these walls. The investigations are based on an effective model focussing on the colloids alone. Our results demonstrate that the fluid-wall interaction stemming from charged walls has a crucial impact on the fluid's high-density behavior as compared to the case of uncharged walls. In particular, based on an analysis of in-plane bond order parameters we find surface-charge-induced freezing and melting transitions.}

\begin{document}

\maketitle

\section{Introduction}
The ordering of charged nanoparticles (colloids) at charged surfaces and in confining geometries is a topic receiving continuous and intense attention since decades. Such systems occur in a wealth of contexts such as in material science (e.g., as candidates for novel self-assembled structures \cite{Vermolen09,Bartlett05}), in microfluidics involving charged channels \cite{Fischer06}, in biological contexts (e.g., proteins at charged lipid membranes \cite{Partha05,Yi08}) and in the physics of dusty plasmas \cite{Klumov07,Fortov05,Messina03}. Moreover, charged colloids have turned out to be excellent model systems with tunable (screened electrostatic) interactions, which are ideally suited to study confinement-related equilibrium and non-equilibrium phenomena such as freezing in unusual crystal structures \cite{Messina03,Oguz09,Mazars08,Grandner08,Fontecha07,Loewen08,Nygaard09}, oscillatory depletion forces \cite{Bechinger99,Klapp08}, and shear-induced distortions of colloidal films \cite{Cohen04,Wu09}.

Despite intense research both from the experimental and from the theoretical (statistical-mechanical) side, we are still far away from a full understanding of the interplay of macroion charges, counterions, salt and surface charges and their precise role for the physics of confined charged suspensions. One prominent, long-standing problem is the occurrence of attractive forces between like-charged colloids between glass surfaces \cite{Rouzina96,Larsen97,Goulding99,Baumgartl06,Allahyarov99,Angelini03,Polin07}, a phenomenon which has been attributed to strong counterion correlations (typical for multivalent counterions) or counterion depletion \cite{Allahyarov99}. Another topic, which is also relevant for moderately coupled systems, concerns the impact of wall charges on the microscopic concentration profiles \cite{Grandner09}. Indeed, recent experimental studies \cite{Nygaard09,Sata08,Sata09} based on x-ray diffraction methods have reported novel effects such as salt-induced trapping of colloidal monolayers \cite{Sata08} and surface-charge-induced stabilization of layers with respect to buckling transitions \cite{Sata09}.

In the present letter we report computer simulation results on yet another phenomenon, that is, the role of surface charges on the {\em lateral} ordering of a charged colloidal suspension between two planar, like-charged walls. In an earlier, combined theoretical-experimental study \cite{Grandner09} we have already shown that such wall charges (which stem from the counterion dissociation into the adjacent fluid) have a crucial impact on the effective fluid-solid interaction and the resulting oscillatory structural forces. Moreover, the theoretically predicted effects for low densities coincide with experimental results \cite{Grandner09}.

Here we go one step further and investigate the confined fluid's crystallization properties. As in our previous studies \cite{Grandner08,Klapp08,Grandner09,Klapp07,Klapp08b,Klapp10a} we employ grand canonical Monte Carlo (GCMC) simulations based on an effective model of the confined colloidal suspension. Our present results demonstrate, for the first time, that surface charges do not only affect the vertical ordering \cite{Grandner09}, but also the (capillary) freezing transitions of the system. In particular, variation of the surface charge can induce freezing and melting transitions, and it can even change the character of the solid phases.

\section{Model system}
\label{model}
Following our previous work \cite{Klapp07,Klapp08,Klapp08b,Grandner08,Grandner09,Klapp10a} we use a model based on Derjaguin-Landau-Verwey-Overbeek (DLVO) theory \cite{Verwey1948}, where the interaction between the charged colloidal spheres suspended in a solution is described on an effective level assuming relatively small macroion density and charge. All other contributions from the dissociated (monovalent) counterions and the salt ions are treated implicitly. The resulting interaction potential reads
\begin{equation}
 \label{equ_u_DLVO}
 u_\mathrm{DLVO}(r) = W\,\frac{\exp(-\kappa r)}{r}
\end{equation}
with the prefactor
\begin{equation}
 \label{W}
 W=\frac{(\tilde{Z} e_0)^2}{4\pi\epsilon_0\epsilon} \exp(\kappa\sigma).
\end{equation}
In eq.~(\ref{W}), $\sigma$ is the diameter of the particles, $e_0$ is the elementary charge, $\epsilon_0$ and $\epsilon$ are the permittivities of the vacuum and the solvent, respectively, and $\tilde{Z}=Z/(1+\kappa\sigma/2)$ is an effective valency assuming that the macroion density is small compared to the salt concentration. The valency of the particles is $Z$. The inverse Debye screening length $\kappa$  appearing in eq.~(\ref{equ_u_DLVO}) is defined as
\begin{equation}
 \label{equ_kappa}
 \kappa = \sqrt{\frac{e_0^2}{\epsilon_0\epsilon k_\mathrm{B} T}(Z\rho + 2I N_{\mathrm{A}})},
\end{equation}
with macroion density $\rho$, ionic strength $I$, Avogadro's constant $N_\mathrm{A}$, Boltzmann constant $\kB$, and temperature $T$. Thus, we do not take into account any effect of the wall counterions on the effective fluid-fluid interaction potential. A justification is given in the discussion of fig.~\ref{fig_corr}.

As in our previous work \cite{Klapp07,Klapp08,Klapp08b,Grandner08,Grandner09,Klapp10a} we supplement the DLVO interaction by a soft-sphere potential $u_\mathrm{SS}(r) = 4\epsilon_\mathrm{SS}(\sigma/r)^{12}$.

\begin{figure}
 \onefigure[width=0.46\textwidth]{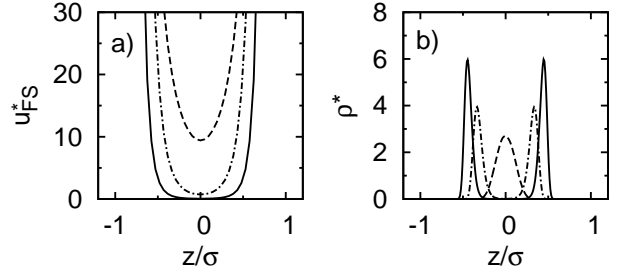}
 \caption{(a) Fluid-solid interaction potential $u_\mathrm{FS}^*(z)$ and (b) local density profiles $\rho^*(z)$ of the macroion distribution for $L_\mathrm{z}=2.4\sigma$ and $\rho_\mathrm{b}^*=0.61$ at surface potentials $|\psi_\mathrm{S}|=0$ (solid lines), $|\psi_\mathrm{S}|=80mV$ (dashed lines), and $|\psi_\mathrm{S}|=160mV$ (dot-dashed lines). The confining walls are located at $z=-1.2\sigma$ and $1.2\sigma$.}
 \label{fig_pot_locdens}
\end{figure}

The fluid-solid interaction between macroions and walls based on linearized Poisson-Boltzmann (PB) theory for small Debye screening lengths and the linear superposition approximation (LSA) reads \cite{Gregory75,Bhattacharjee98}
\begin{equation}
 \label{equ_u_lsa}
 u_\mathrm{FS}^\mathrm{LSA}(z) = 64\pi\epsilon_0\epsilon\gamma_\mathrm{F}\gamma_\mathrm{S}
                    \frac{\sigma}{2}\left(\frac{\kB T}{e_0}\right)^2 
                    \exp\left(-\kappa_\mathrm{W}(z)(z-\frac{\sigma}{2})\right),
\end{equation}
where $\gamma_\mathrm{F/S} = \tanh\left(e_0\psi_\mathrm{F/S}/4\kB T\right)$, and $\psi_\mathrm{F/S}$ is the surface potential of the fluid particles (F) and the solid walls (S). Contrary to \cite{Bhattacharjee98}, however, the screening parameter $\kappa_\mathrm{W}(z)$ depends in our model on the $z$-coordinate, reflecting the inhomogeneous distribution of counterions released by the walls. Specifically, using the exact PB expression for the counterion profile \cite{Allahyarov99} in the dilute limit, and performing an averaging procedure, our expression for the space-dependent wall screening parameter reads \cite{Grandner09} 
\begin{equation}
 \label{equ_kappaW}
 \kappa_\mathrm{W}(z) = \sqrt{ \frac{e_0^2}{\epsilon_0\epsilon\kB T} 
                        \left( Z\rho + 2I N_\mathrm{A} 
                        + \frac{\left|\sigma_\mathrm{S}\right|}{e_0 z}\right) }
\end{equation}
with the surface charge density $\sigma_\mathrm{S}$. The fluid-solid interaction is supplemented by a soft-wall repulsion $u_\mathrm{FS}^\mathrm{SW}(z) = (4/5)\pi\epsilon_\mathrm{w}(\sigma/z)^9$. The resulting fluid-solid potential $u_\mathrm{FS}(z)=\kB T u_\mathrm{FS}^*(z)=u_\mathrm{FS}^\mathrm{SW}(z)+u_\mathrm{FS}^\mathrm{LSA}(z)$ shown in fig.~\ref{fig_pot_locdens}a reveals a non-monotonic behavior as function of $\sigma_\mathrm{S}$ resulting from a competition between Coulomb repulsion and screening \cite{Grandner09}.
The same qualitative behavior of $u_\mathrm{FS}$ is observed with a homogeneous ansatz for $\kappa_\mathrm{W}$ \cite{Allahyarov99,Loewen08}, as discussed in detail in \cite{Grandner09}. We note, that our ansatz in eq.~(\ref{equ_kappaW}) yields a dependence of resulting surface forces on the wall charge consistent with the experiment. Indeed, the data in fig.~\ref{fig_pot_locdens}a are based on realistic values of $\psi_\mathrm{S}$ such as silica ($|\psi_\mathrm{S}|=80mV$) and mica ($|\psi_\mathrm{S}|=160mV$) which are typical materials for experiments \cite{Grandner09}. For increasing $\sigma_\mathrm{S}$ the repulsion range of the surfaces increases and the pore is effectively narrowed. Above a certain charge the wall screening begins to dominate the Coulomb repulsion and yields an effective broadening. However, the effective pore width for the highest surface charge considered (mica) is still smaller compared to the uncharged case. The non-monotonic behavior of $u_\mathrm{FS}$ has a strong impact on the particle layering of the macroions as shown in fig.~\ref{fig_pot_locdens}b where we present density profiles $\rho_\mathrm{z}$ \cite{Grandner08}.
We note that these effects remain when $\kappa$ (eq.~(\ref{equ_kappa})) appearing in the fluid-fluid interaction is replaced by a screening parameter involving the surface charge explicitly, as we show in fig.~\ref{fig_corr}.
%
Our model only contains pure repulsive interactions. As motivated in \cite{Grandner09} like-charge attraction \cite{Rouzina96,Larsen97, Allahyarov99, Goulding99, Angelini03, Baumgartl06,Polin07} between the macroions is {\em not} expected in the parameter range of interest. 

Parameters are set in order to mimic the conditions of Colloidal-Probe Atomic-Force-Microscope experiments involving dilute silica suspensions with low salt concentrations \cite{Klapp07,Grandner09,Klapp08b,Klapp10a}. Resulting values of the coupling and screening parameters are $W^*=(\kB T)^{-1}W\approx80-90$ and $\kappa^*=\kappa\sigma\approx2.5-2.7$ \cite{Grandner08}.
In-plane correlation functions $g_{||}(r_{||}) = \av{N_\mathrm{l}(r_{||},\Delta r)}/(N_\mathrm{l}\rho_\mathrm{l}2\pi r_{||}\Delta r)$ for a certain layer $l$ with $N_\mathrm{l}$ particles within a small interval $\Delta r$ for the in-plane particle separation $r_{||} = |\vec r_{kj}| = \sqrt{(x_{kj}^2 + y_{kj}^2)}$ of particles $k$ and $j$, and the particle area density $\rho_\mathrm{l}$ are extracted from the simulations.
Another key quantity is the local bond order parameter $\psi_\mathrm{n} = \av{1/N_\mathrm{layer}\sum_{k=1}^{N_\mathrm{layer}} 1/N^\mathrm{b}_k \left| \sum\nolimits_{j=1}^{N^\mathrm{b}_k} \exp(\mathrm{i} n \theta_{kj}) \right|}$ with the angle $\theta_{kj}$ enclosing the bond vector $\vec r_{kj}$  and an arbitrary in-plane axis. It is a meassure for the local in-plane order of a certain layer parallel to the confining walls with $N_\mathrm{layer}$ particles \cite{Grandner08}. In order to find the $N^\mathrm{b}_k$ next neighbors of a particle we employed a Delaunay triangulation \cite{NumRecipes}. Crystal-like ordering is signalled by values of $\psi_\mathrm{n}$ larger than $0.6$ as motivated by the results in fig.~\ref{fig_psiS_psi8_psi6}.

\section{Results and Discussion}
\label{results}


We have performed GCMC simulations with confined systems containing about 500, 1000 and 2000 particles. After equilibration of the systems using $50,000$ - $500,000$ MC steps we collected averages of relevant physical quantities with $100,000$ - $1,000,000$ MC steps. Our present investigations focus on nanoconfined suspensions characterized by surface separations $L_\mathrm{z}$ yielding one- or two layer systems. 

\begin{figure}
 \onefigure[width=0.49\textwidth]{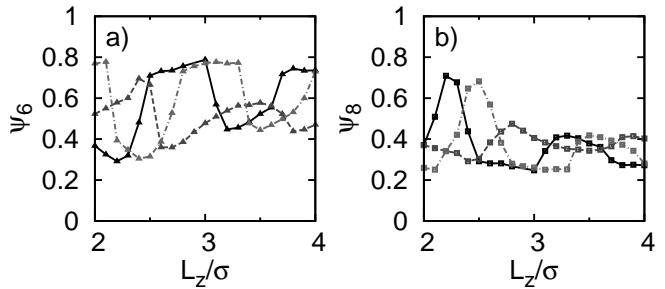}
 \caption{Bond order parameters (a) $\psi_6$ and (b) $\psi_8$ as function of the wall separation $L_\mathrm{z}$ at $\rho_\mathrm{b}^*=0.61$ for uncharged (solid line), silica (dashed line), and mica (dot-dashed line) walls. The triangles and squares are the simulation data, and the lines are a guide to the eye.}
 \label{fig_Lz_psi6_psi8}
\end{figure}

A first impression of the role of both, the value of $L_\mathrm{z}$ and the surface charge, is given in fig.~\ref{fig_Lz_psi6_psi8}, where we plotted bond order parameters as function of $L_\mathrm{z}$ for $\rho_\mathrm{b}^*=0.61$. Specifically, we consider $\psi_6$ and $\psi_8$ which are sensitive to hexagonal and squared order, respectively. We used $\psi_8$ instead of $\psi_4$ since the next neighbors were determined via Delaunay triangulation where on average the particles have six next neighbors. Therefore, in contrast to $\psi_8$ the parameter $\psi_4$ cannot become $1.0$ even in a perfect squared lattice. As known from previous work \cite{Grandner08} the bond order parameters exhibit an alternating behavior as function of $L_\mathrm{z}$ stemming from a competition between layering and in-plane ordering. As seen in fig.~\ref{fig_Lz_psi6_psi8} this behavior is significantly affected by the wall potential. On one hand the phase of the alternation is shifted forwards larger separations when the uncharged walls are replaced by silica walls since the interaction range of the wall potential increases (see fig.~\ref{fig_pot_locdens}). On the other hand the height of the maxima of $\psi_6$ and $\psi_8$ becomes smaller because the layers are softened \cite{Grandner09}. For larger surface charges up to mica ($|\psi_\mathrm{S}| \gtrsim 100mV$) the opposite behavior occurs. The oscillations are shifted back and the maxima are enhanced. In particular, looking at a certain separation such as, e.~g., $L_\mathrm{z}=2.4\sigma$ the parameter $\psi_6$ jumps from $0.5$ (uncharged) to $0.7$ (silica) and back to $0.3$ (mica). Similar behavior is observed for $\psi_8$. Mica surfaces exhibit a bond order larger than $0.6$ whereas the other wall types yield much smaller values.

\begin{figure}
 \onefigure[width=0.4\textwidth]{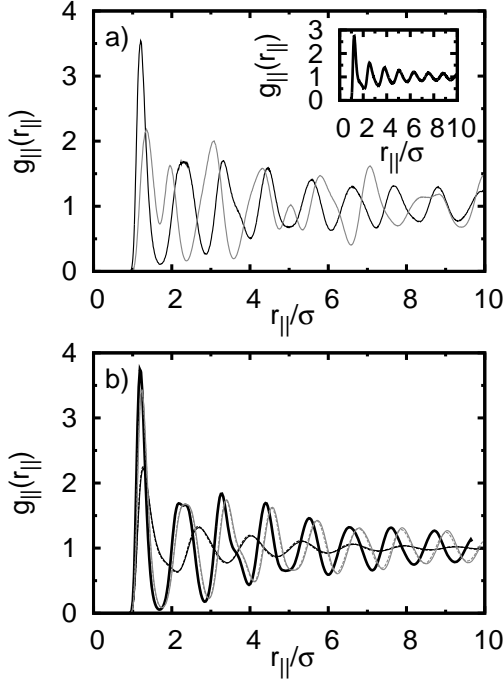}
 \caption{(a) In-plane correlation functions for $L_z=2.4\sigma$ using the wall types uncharged (inset, thick black solid line), silica (thin black solid line) and mica (thin grey solid line). (b) In-plane correlation functions for $L_z=2.9\sigma$ with wall types as in (a). Included are results involving the wall-counterion contribution in the fluid-fluid interaction for silica (thin black dashed line) and mica (thin grey dashed line).}
 \label{fig_corr}
\end{figure}

This charge-induced ordering and disordering is also visible from the in-plane correlation functions $g_{||}(r_{||})$. In fig.~\ref{fig_corr}a correlation functions are plotted for the three different wall types, $L_z=2.4\sigma$ and $\rho_\mathrm{b}^*=0.61$.
For uncharged surfaces the correlations behave still fluid-like. On the other hand, silica and mica walls both induce crystal-like correlation functions reflecting hexagonal and squared order for silica and mica, respectively. Data for $L_\mathrm{z}=2.9\sigma$ are shown in fig.~\ref{fig_corr}b. In this case, uncharged and mica walls exhibit crystal-like translational correlations whereas silica walls correspond to fluid-like behavior. Both results are consistent with the charge-induced ordering of the particles seen in fig.~\ref{fig_Lz_psi6_psi8}.

At this point, we briefly discuss the influence of the additional wall counterions on the effective fluid-fluid interaction. Such an approach was suggested in \cite{Allahyarov99}, where the screening parameter of the DLVO fluid-fluid interaction includes a counterion contribution $\tilde\rho_\mathrm{cw}$ stemming from the surface charge defined as an average of the microscopic (PB) profile, that is,
\begin{equation}
 \label{equ_kappa_wcFF}
 \tilde\kappa = \tilde\kappa(z_1,z_2) = \sqrt{\frac{e_0^2}{\epsilon_0\epsilon k_\mathrm{B} T}(Z\rho + 2I N_{\mathrm{A}} + \tilde\rho_\mathrm{cw}(z_1,z_2))}.
\end{equation} 
We tested the importance of this modification of the fluid-fluid interaction by calculating corresponding in-plane correlations at $L_\mathrm{z}=2.9\sigma$. The results in fig.~\ref{fig_corr}b, however, show that the effect is marginal.

\begin{figure}
 \onefigure[width=0.46\textwidth]{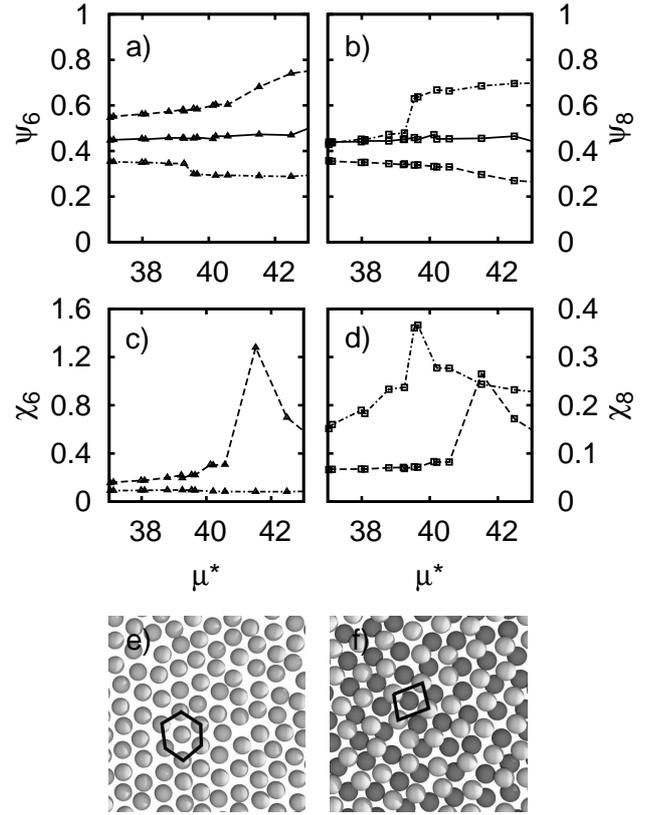}
 \caption{Bond angle order parameters (a) $\psi_6$ and (b) $\psi_8$ with corresponding susceptibilities (c) $\chi_6$ and (d) $\chi_8$ as function of $\mu^*$ for uncharged (solid line), silica (dashed line), and mica (dot-dashed) walls with a separation of $L_\mathrm{z}=2.4\sigma$. The triangles and the squares are simulation results, whereas the lines are a guide to the eye. Corresponding top-view snapshots for $\mu^*=41.52$ and (e) silica walls (hexagonal phase) and (f) mica walls (squared phase).}
 \label{fig_Lz2.4psi}
\end{figure}

We now consider the impact of wall charges on the systems ordering behavior as function of the chemical potential (and thus, the density), focussing
on the wall separation $L_\mathrm{z}=2.4\sigma$. Figures \ref{fig_Lz2.4psi}a and \ref{fig_Lz2.4psi}b show corresponding results for $\psi_6$ and $\psi_8$ for uncharged, silica ($|\psi_\mathrm{S}|=80mV$), and mica walls ($|\psi_\mathrm{S}|=160mV$).
Whereas $\psi_6$ for uncharged and mica walls is less than $0.5$, silica walls induce a significant hexagonal ordering which becomes particularly pronounced for $\mu^*>40.5$ ($\rho_\mathrm{b}^*>0.59$). Indeed, at the largest $\mu^*$ considered the order parameter $\psi_6$ has values of about $0.75$ consistent with the clear hexagonal ordering seen in fig.~\ref{fig_Lz2.4psi}e. Another type of charge-induced ordering occurs for mica walls. This is clearly visible from fig.~\ref{fig_Lz2.4psi}b, which reveals a marked increase of $\psi_8$ for mica walls, whereas corresponding order parameters remain constant or even decrease for uncharged and silica walls,
respectively.
Summarizing, for large enough $\mu^*$ an increase of the surface potential from $|\psi_\mathrm{S}|=0$ (uncharged) to $|\psi_\mathrm{S}|=80mV$ (silica) and $|\psi_\mathrm{S}|=160mV$ (mica) yields a change of the in-plane order from non-ordered to hexagonal and finally to squared. Moreover, the onset of translational ordering is slightly shifted towards smaller chemical potentials when silica is replaced by mica (uncharged walls do not order at all in the density range considered).

To confirm the picture that the charged walls lead to true crystallization phase transitions we have computed fluctuations (susceptibilities) of the bond-order parameters defined as $\chi_\mathrm{n}=N(\av{\psi_\mathrm{n}^2}-\av{\psi_\mathrm{n}}^2)$. These fluctuations are expected to increase near a phase transition, i.~e., at the phase transition $\chi_\mathrm{n}$ should exhibit a maximum \cite{Binder87}. Figure \ref{fig_Lz2.4psi}c contains the susceptibility $\chi_6$ corresponding to the data for $\psi_6$ in fig.~\ref{fig_Lz2.4psi}a. This quantity possesses for silica a maximum around $\mu^*=41.5$ ($\rho_\mathrm{b}^*=0.61$) which can be identified with the onset of hexagonal ordering. As expected $\chi_6$ remains small for mica. Considering squared structures $\chi_8$ exhibits a maximum at $\mu^*=39.5$ ($\rho_\mathrm{b}^*=0.58$) in fig.~\ref{fig_Lz2.4psi}d which is consistent with the strong increase of $\psi_8$ in fig.~\ref{fig_Lz2.4psi}b. Although the bond order parameter $\psi_8$ remains small for silica walls it also shows a maximum in $\chi_8$. This non-expected maximum for silica in fig.~\ref{fig_Lz2.4psi}d is related to strong bond order fluctuations when the hexagonal order in the system emerges (see $\psi_6$ and $\chi_6$).

\begin{figure}
 \onefigure[width=0.49\textwidth]{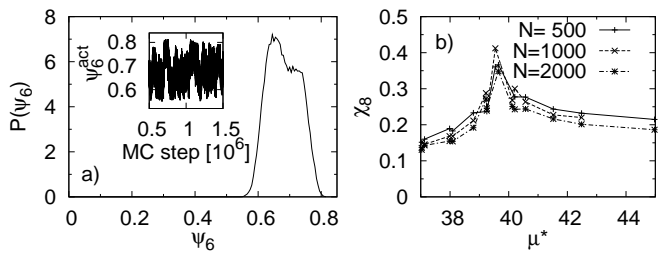}
 \caption{(a) Distribution $P(\psi_6)$ of the bond angle order $\psi_6$ for $L_\mathrm{z}=2.4\sigma$ and $\rho_\mathrm{b}^*=0.61$ ($\mu^*=41.52$) using silica walls. The inset shows the actual non-averaged values of $\psi_6$ during the MC simulation in the averaging interval. (b) System size dependence of the bond order fluctuations $\chi_8$ as function of the chemical potential $\mu^*$ for mica surfaces with $L_\mathrm{z}=2.4\sigma$ and different particle numbers $N=500,1000,2000$. The symbols are the simulation data and the lines are a guide to the eye.}
 \label{fig_psi6_histo}
\end{figure}

We note in this context that the results for the susceptibilities close to the phase transition are plagued by the fact that the simulated system ''jumps'' between two (coexisting) phases, despite the large free energy barrier separating fluid and solid phases. Indeed, most of the data points in figs.~\ref{fig_Lz2.4psi}c and \ref{fig_Lz2.4psi}d do not contain any jumps, an exception being the data corresponding to $\mu^*=41.52$ and $\mu^*=42.48$ for silica walls with large $\chi_6$ and $\chi_8$. For $\mu^*=41.52$ we show in fig.~\ref{fig_psi6_histo}a results for the order parameter distribution $P(\psi_6)$. The latter reflects a double-peak structure typical for a first-order transition near coexistence \cite{Wang01,FrenkelSmit}. The peaks overlap due to the similarity of the order parameter in the two phases. The corresponding behavior of the instantaneous value of $\psi_6$ as function of MC ''time'' clearly shows not only fluctuations on short scale, but also jumps between the two phases on a longer time scale.

The first-order character of the observed phase transitions is also indicated by the behavior of the susceptibilities as functions of the system size. Specifically, for a second-order phase transition the susceptibility at phase coexistence is supposed to diverge in the thermodynamic limit whereas for a first-order transition this maximum does not depend on $N$ \cite{Binder87}. As an example, we consider in fig.~\ref{fig_psi6_histo}b results for the emerging squared phase between mica surfaces (see figs.~\ref{fig_Lz2.4psi}b and \ref{fig_Lz2.4psi}d). Computing $\chi_8$ for the particle numbers $N=500,1000,2000$ we see that the maximum remains more or less the same (fig.~\ref{fig_psi6_histo}b). The small deviations are supposed to be due to statistical errors. We conclude that the transitions found in figs.~\ref{fig_Lz2.4psi}b and \ref{fig_Lz2.4psi}d are of first order.

\begin{figure}
 \onefigure[width=0.49\textwidth]{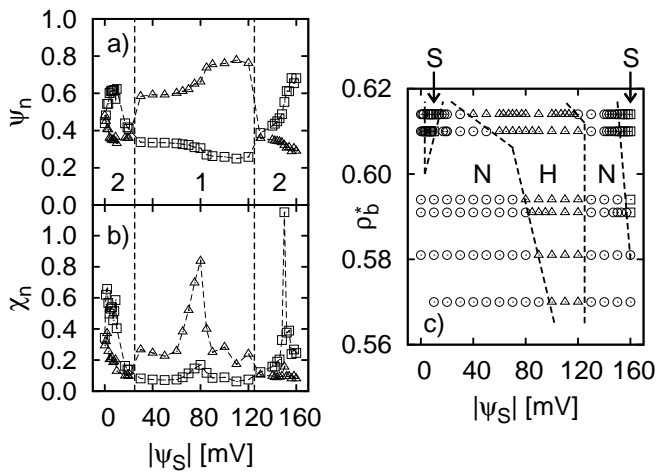}
 \caption{Influence of surface charges (surface potential $\psi_\mathrm{S}$) on the in-plane order for $L_\mathrm{z}=2.4\sigma$. (a) Bond order parameters $\psi_8$ (squares) and $\psi_6$ (triangles) as function of $\psi_\mathrm{S}$. The number of layers are indicated by the vertical dashed lines and the numbers. (b) Corresponding bond order susceptibilities $\chi_8$ (squares) and $\chi_6$ (triangles). (c) Phase diagram as function of $\psi_\mathrm{S}$ and bulk density $\rho_\mathrm{b}^*$ showing squared phases (squares, 'S') with $\psi_8\geq0.6$, hexagonal phases (triangles, 'H') with $\psi_6\geq0.6$, and non-ordered phases (circles, 'N'). The dashed lines are a schematic representation of the phase boundaries. }
 \label{fig_psiS_psi8_psi6}
\end{figure}

So far we have concentrated on density-driven crystallization transitions at two particular surface charges corresponding to silica and mica walls, respectively. However, as we will demonstrate below, such transitions also occur when varying the surface charge {\em continuously} at fixed chemical potential (corresponding to a bulk density of $\rho_\mathrm{b}^*=0.61$). Our results for $\psi_\mathrm{n}$ plotted in fig.~\ref{fig_psiS_psi8_psi6}a reveal two major effects of such an ''experiment'', the first concerning the number of layers observed at given $\psi_\mathrm{S}$. Indeed, increasing $|\psi_\mathrm{S}|$ from zero one first finds a rather abrupt change (as revealed by density profiles) from a two- to a one-layer system, reflecting the effective narrowing of the pore due to increasing Coulomb repulsion (see also fig.~\ref{fig_pot_locdens}). Further charging then yields a second transformation back to a two-layer system, consistent with the fact that the range of repulsion decreases with increasing $|\psi_\mathrm{S}|$ for surface charges beyond $|\psi_\mathrm{S}|\approx100mV$. The charge-induced changes of the number of layers are accompanied by sudden changes of the order parameter values, as one might expect. More significantly, however, is the fact that the order parameters exhibit further sudden changes {\em within} the ranges of constant layer number, accompanied by sharp maxima of the corresponding susceptibilities (see fig.~\ref{fig_psiS_psi8_psi6}b). Specifically, increasing $|\psi_\mathrm{S}|$ from zero one detects a first transition (within the two-layer regime) at $|\psi_\mathrm{S}|\approx10mV$, where the fluid crystallizes into a state with square-like order. For slightly larger $|\psi_\mathrm{S}|$, the system melts again into a fluid state. Similarly, there is a fluid-to-hexagonal transition at $|\psi_\mathrm{S}|\approx80mV$ and a fluid-to-square transition at $|\psi_\mathrm{S}|\approx160mV$ accompanied by large bond order susceptibilities. Note, that the transitions between one- and two-layer systems are not symmetric, i.~e., the two-layer system ''jumps'' for small $|\psi_\mathrm{S}|$ into an one-layered fluid phase, whereas for high $|\psi_\mathrm{S}|$ it ''jumps'' into a one-layered hexagonal structure. Taken altogether, the results in fig.~\ref{fig_psiS_psi8_psi6}a confirm the idea that variation of the surface charge can induce freezing and melting transitions. 

Finally, we give in fig.~\ref{fig_psiS_psi8_psi6}c an overview of the system's behavior at the exemplary wall separation $L_\mathrm{z}=2.4\sigma$ including results at various densities in the range $0.57 \leq \rho_\mathrm{b}^* \leq 0.614$. The resulting ''phase diagram'' in the $\psi_\mathrm{S}$-$\rho_\mathrm{b}^*$ plane displays non-ordered (N), squared (S) and hexagonal (H) phases. The diagram also contains the results of figs.~\ref{fig_psiS_psi8_psi6}a and \ref{fig_psiS_psi8_psi6}b for $\rho_\mathrm{b}^*=0.61$. 
The observed phase sequences between non-ordered and ordered phases including reentrance of the H phase are a result of the varying shape of $u_\mathrm{FS}$ with $\psi_\mathrm{S}$ (see fig.~\ref{fig_pot_locdens}a) combined with the value of $L_\mathrm{z}$ considered.
The bond order threshold for distinguishing non-ordered and ordered phases is set to $0.6$, i.~e., $\psi_6\geq0.6$ is considered as hexagonal and $\psi_8\geq0.6$ as squared. This criterion is motivated by the fact that the freezing transitions indicated by the peaks in the bond order susceptibilities $\chi_\mathrm{n}$ in fig.~\ref{fig_psiS_psi8_psi6}b correspond to values $\psi_\mathrm{n}\approx0.6$. However, the choice of the threshold does not change the qualitative behavior of the phase diagram.
We note that at the parameters investigated we did not find any other complex, such as buckled, structures which have been observed in hard sphere systems \cite{Fortini06} and ground-state Coulomb systems \cite{Oguz09b}. One reason for the absence of buckling in our system might be the rather small value of $\kappa$ and the finite temperature. Indeed, according to a recent ground-state study \cite{Oguz09}, complex structures typically occur at larger screening parameters. Furthermore, a finite temperature MC study of Yukawa bilayers \cite{Mazars08} found different (staggered) phases without buckling. In our system the region between one- and two-layer systems (see fig.~\ref{fig_psiS_psi8_psi6}c), where buckling might emerge, is characterized by a fluid-like order with small bond order parameters.

\section{Conclusion}
\label{conclusion}

In summary, our computer simulation results demonstrate that surface charges have a crucial impact on the lateral ordering of charged colloidal suspensions confined to nanoscopic films. The observed effects originate from the impact of the wall counterions on the fluid-wall interaction, whereas the details of the fluid-fluid interaction play only a minor role. Most spectacularly, we find that charged walls support crystalline, hexagonal- and square-ordered phases (depending on the actual surface potential) under conditions where the corresponding suspension between uncharged walls (and its bulk counterpart) is still fluid-like. Changing continuously the wall charge at fixed wall separation and density we find both, layering transitions and first-order, in-plane freezing and melting transitions including reentrant behavior.
The exemplary cases studied in the present work indicate that charges on confining substrates add a wealth of structure formation phenomena to the already rich and complex phase diagram of confined charged colloids or dusty plasmas.

Of course, future experimental work is needed to confirm these theoretical predictions, one first indication being the unusual shape of the oscillatory surface forces characterizing dense suspensions of silica nanoparticles between (charged) silica surfaces \cite{Klapp08}. Moreover, it would be very interesting and also important from a material science point of view \cite{Vermolen09,Bartlett05} to extend the investigations towards the case of binary colloidal mixtures of differently sized and charged particles. Work in this direction is under way \cite{Klapp10a}.

We thank the Deutsche Forschungsgemeinschaft for financial support via grant KL1215/6.



\begin{thebibliography}{99}

\bibitem{Vermolen09} Vermolen E.~C.~M., Kuijk A., Filion L.~C., Hermes M., Thijssen J.~H.~J., Dijkstra M., and van Blaaderen A., {\it Proc.~Nat.~Acad.~Sci.}, {\bf 106} (2009) 16063.
\bibitem{Bartlett05} Bartlett P.~and Campbell A.~I., {\it Phys.~Rev.~Lett.}, {\bf 95} (2005) 128302.

\bibitem{Fischer06} Helseth L.~E., Wen H.~Z., and Fischer T.~M., {\it J.~Appl.~Phys.}, {\bf 99} (2006) 024909.

\bibitem{Partha05} Parthasarathy R., Cripe P.~A., and Groves J.~T., {\it Phys.~Rev.~Lett.}, {\bf 95} (2005) 048101.
\bibitem{Yi08} Yi M., Nymeyer H., and Zhou H.-X., {\it Phys.~Rev.~Lett.}, {\bf 101} (2008) 038103.

\bibitem{Klumov07} Klumov B.~A. and Morfill G.~E., {\it JETP Letters}, {\bf 85} (2007) 498. 
\bibitem{Fortov05} Fortov V.~E., Ivlev A.~V., Khrapak S.~A., Khrapak A.~G., Morfill G.~E., {\it Phys.~Rep.}, {\bf 421} (2005) 1.
\bibitem{Messina03} Messina R. and L\"owen H., {\it Phys.~Rev.~Lett.}, {\bf 91} (2003) 146101.

\bibitem{Oguz09} O\u{g}uz E.~C., Messina R. and L\"owen H., {\it Europhys.~Lett.}, {\bf 86} (2009) 28002.
\bibitem{Mazars08} Mazars M., {\it Europhys.~Lett.}, {\bf 84} (2008) 55002.
\bibitem{Grandner08} Grandner S. and Klapp S.~H.~L., {\it J.~Chem.~Phys.}, {\bf 129} (2008) 244703.
\bibitem{Fontecha07} Barreira Fontecha A., Palberg T., and Sch\"ope H.~J., {\it Phys.~Rev.~E}, {\bf 76} 050402(R) (2007).
\bibitem{Loewen08} L\"owen H., H\"artel A., Barreira-Fontecha A., Sch\"ope H.~J., Allahyarov E., and Palberg T., {\it J.~Phys.: Condens.~Matter}, {\bf 20} (2008) 404221.
\bibitem{Nygaard09} Nyg\aa{}rd K., Satapathy D.~K., Buitenhuis J., Perret E., Bunk O., David C., and van der Veen J.~F., {\it Europhys.~Lett.}, {\bf 86} (2009) 66001.

\bibitem{Bechinger99} Bechinger C., Rudhardt D., Leiderer P., Roth R., and Dietrich S., {\it Phys.~Rev.~Lett.}, {\bf 83} (1999) 3960.
\bibitem{Klapp08} Klapp S.~H.~L., Zeng Y., Qu D., and von Klitzing R., {\it Phys.~Rev.~Lett.}, {\bf 100} (2008) 118303.

\bibitem{Cohen04} Cohen I., Mason T.~G., and Weitz D.~A., {\it Phys.~Rev.~Lett.}, {\bf 93} (2004) 046001.
\bibitem{Wu09} Wu Y.~L., Derks D., van Blaaderen A., and Imhof A., {\it Proc.~Nat.~Acad.~Sci.}, {\bf 106} (2009) 10564.

\bibitem{Rouzina96} Rouzina I. and Bloomfield V.~A., {\it J.~Phys.~Chem.}, {\bf 100} (1996) 9977.
\bibitem{Larsen97} Larsen A.~E. and Grier D.~G., {\it Nature}, {\bf 385} (1997) 230.
\bibitem{Goulding99} Goulding D. and Hansen J.-P., {\it Europhys.~Lett.}, {\bf 46} (1999) 407.
\bibitem{Baumgartl06} Baumgartl J., Arauz-Lara J.~L. and Bechinger C., {\it Soft Matter}, {\bf 2} (2006) 631.
\bibitem{Allahyarov99} Allahyarov E., D'Amico I., and L\"owen H., {\it Phys.~Rev.~E}, {\bf 60} (1999) 3199.
\bibitem{Angelini03} Angelini T.~E., Liang H., Wriggers W., and Wong G.~C.~L., {\it Proc.~Nat.~Acad.~Sci.}, {\bf 100} (2003) 8634.  
\bibitem{Polin07} Polin M., Grier D.~G., and Han Y., {\it Phys.~Rev.~E}, {\bf 76} (2007) 041406.

\bibitem{Grandner09} Grandner S., Zeng Y., von Klitzing R., and Klapp S.~H.~L., {\it J.~Chem.~Phys.}, {\bf 131} (2009) 154702.

\bibitem{Sata08} Satapathy D.~K., Bunk O., Jefimovs K., Nyg\aa{}rd K., Guo H., Diaz A., Perret E., Pfeiffer F., David C., Wegdam G.~H., and van der Veen J.~F., {\it Phys.~Rev.~Lett.}, {\bf 101} (2008) 136103.
\bibitem{Sata09} Satapathy D.~K., Nyg\aa{}rd K., Bunk O., Jefimovs K., Perret E., Diaz A., Pfeiffer F., David C.,
and van der Veen J.~F., {\it Europhys.~Lett.}, {\bf 87} (2009) 34001.

\bibitem{Klapp07} Klapp S.~H.~L., Qu D., and von Klitzing R., {\it J.~Phys.~Chem.~B}, {\bf 111} (2007) 1296.
\bibitem{Klapp08b} Klapp S.~H.~L., Grandner S., Zeng Y., and von Klitzing R., {\it J.~Phys.: Condens.~Matter}, {\bf 20} (2008) 494232.
\bibitem{Klapp10a} Klapp S.~H.~L., Grandner S., Zeng Y., and von Klitzing R., {\it Soft Matter}, {\bf 6} (2010) 2330.

\bibitem{Verwey1948} Verwey E.~J.~W. and Overbeek J.~T.~G., \textit{Theory of the Stability of Lyophobic Colloids}, (Elsevier, Amsterdam) 1948.

\bibitem{Gregory75} Gregory J., {\it J.~Coll.~Int.~Sc.}, {\bf 51} (1975) 44.
\bibitem{Bhattacharjee98} Bhattacharjee S., Elimelech M., and Borkovec M., {\it CCACAA}, {\bf 71} (1998) 883.

\bibitem{NumRecipes} Press W.~H., Teukolsky S.~A., Vetterling W.~T., and Flannery B.~P., {\it Numerical Recipes}, 3rd ed. (Cambridge University Press, New York) 2007.
\bibitem{Binder87} Binder K., {\it Rep.~Prog.~Phys.}, {\bf 50} (1987) 783.

\bibitem{Wang01} Wang F. and Landau D.~P., {\it Phys.~Rev.~Lett.}, {\bf 86} (2001) 2050.
\bibitem{FrenkelSmit} Frenkel D. and Smit B., {\it Understanding Molecular Simulation}, (Academic Press, San Diego) 2002.     

\bibitem{Fortini06} Fortini A. and Dijkstra M., {\it J.~Phys.: Condens.~Matter}, {\bf 18} (2006) L371.

\bibitem{Oguz09b} O\u{g}uz E. C., Messina R. and L\"owen H., {\it J.~Phys.: Condens.~Matter}, {\bf 21} (2009) 424110.

\end{thebibliography}
\end{document}